\documentclass[onecolumn,amsmath,amssymb]{qm2008_abs}
\usepackage{graphicx}
\usepackage{dcolumn}
\usepackage{bm}
\begin{document}

\title{{\Large Dissipative or just Nonextensive hydrodynamics?\\
- Nonextensive/Dissipative correspondence -}}

\bigskip
\bigskip
\author{\large Takeshi Osada}
\email{osada@ph.ns.musashi-tech.ac.jp} \affiliation{Theoretical
Physics Lab., Faculty of Knowledge Engineering, Musashi Institute
of Technology, Setagaya-ku, Tokyo 158-8557, Japan}
\bigskip
\author{\large Grzegorz Wilk}
\email{wilk@fuw.edu.pl} \affiliation{The Andrzej So\l tan
Institute for Nuclear Studies, Theoretical Physics Department,
Ho\.za 69; 00-681 Warsaw, Poland}
\bigskip
\bigskip

\begin{abstract}
\leftskip1.0cm \rightskip1.0cm We argue that there is
correspondence between the perfect nonextensive hydrodynamics and
the usual dissipative hydro\-dy\-na\-mics, which we call {\it
nonextensive/dissipative correspondence (NexDC)}. It leads to
simple expression for dissipative entropy current and allows for
predictions for the ratio of bulk and shear viscosities to entropy
density, $\zeta/s$ and $\eta/s$.
\end{abstract}

\maketitle

\begin{figure}[hbt!]
\begin{center}
\includegraphics[width=7.4cm]{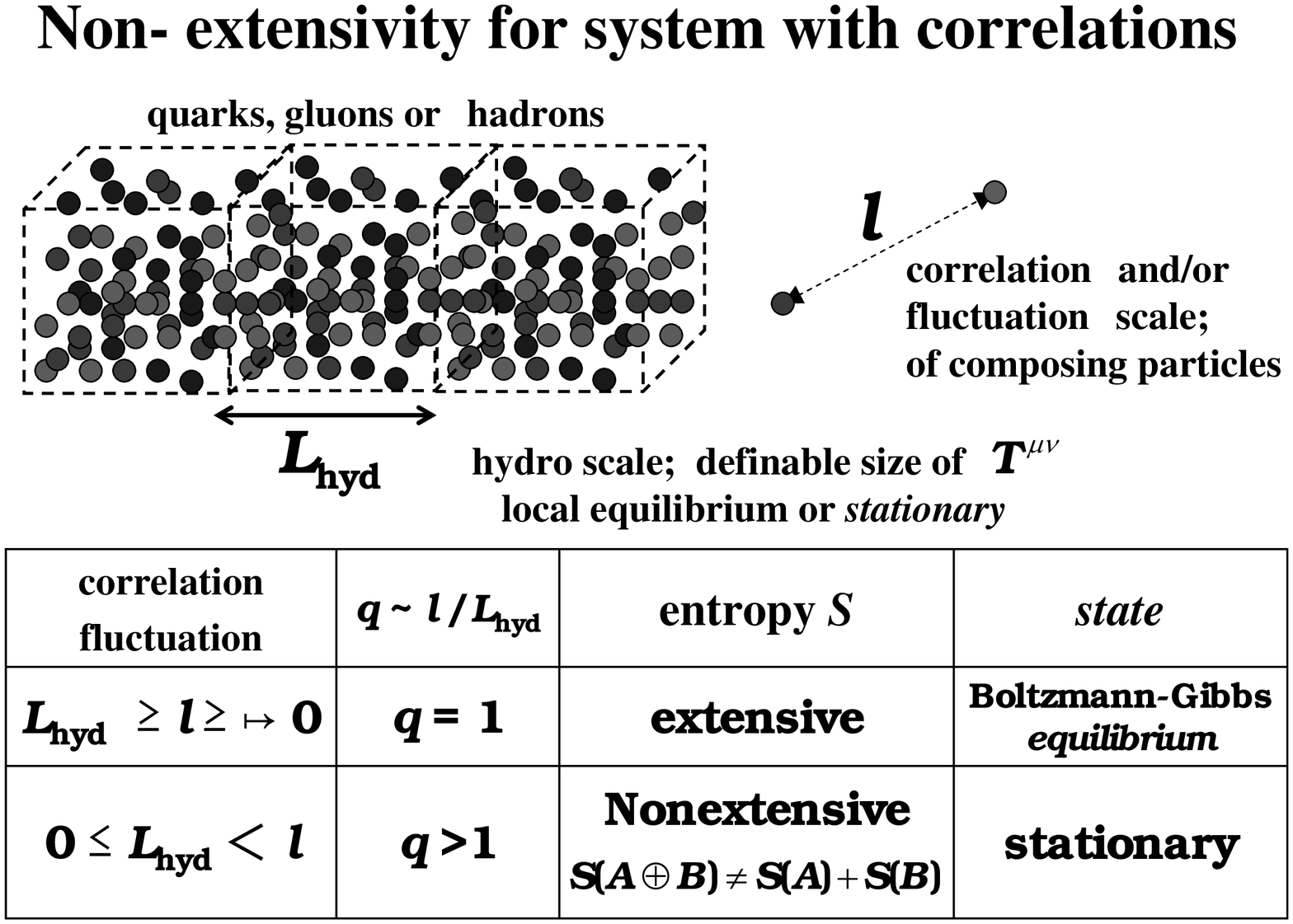}
\includegraphics[width=7.4cm]{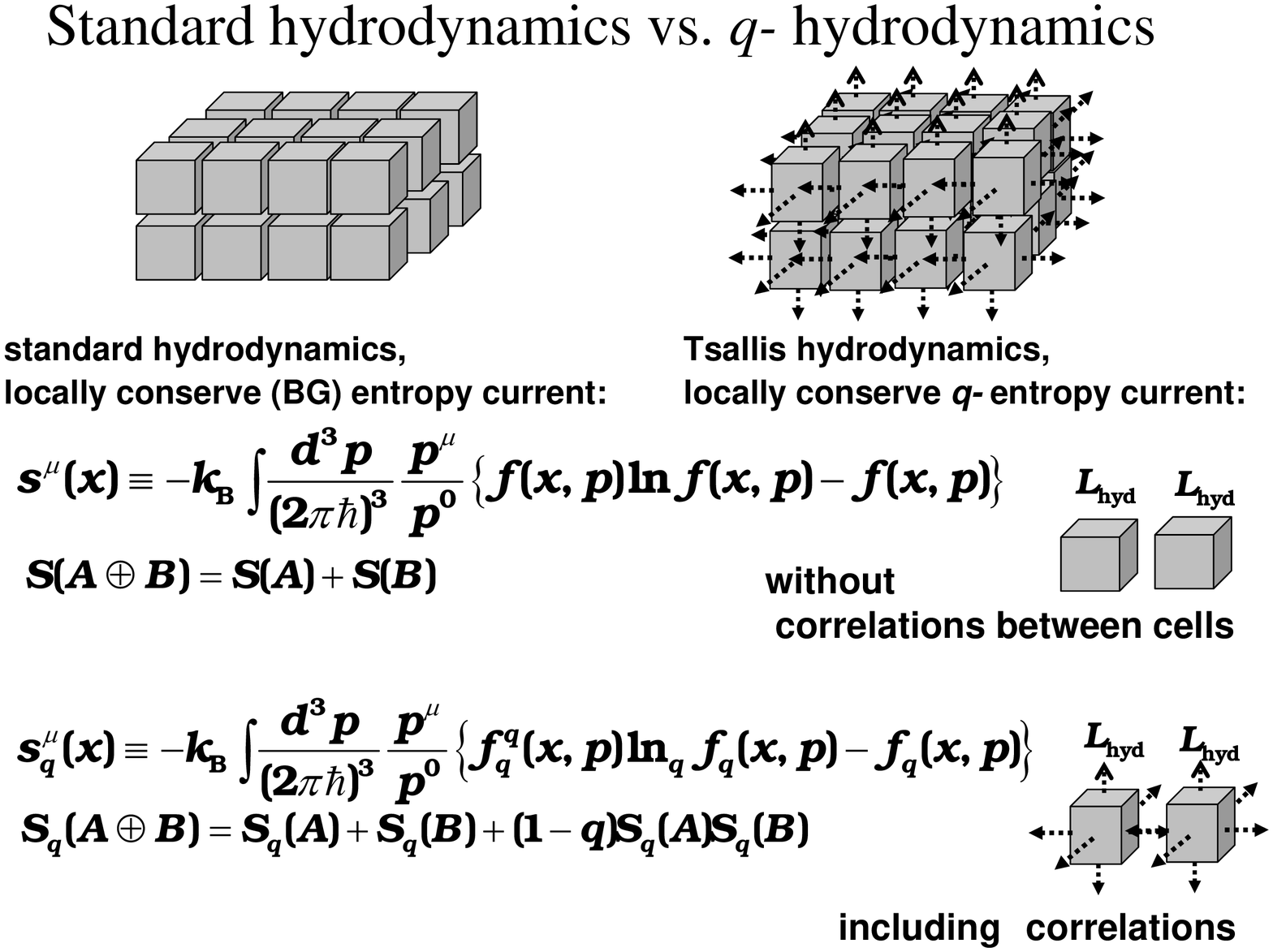}
\caption{Visualization of ideas behind  the
$q$-hydrodynamics.} \label{Fig12}
\end{center}
\end{figure}

Recently there is renewed interest in dissipative hydrodynamical
models \cite{IsraelAnnPhys118,MurongaPRC69,
Tsumura2007,HiscockPhysRevD31, KoidePRC75} prompted by the success
of perfect hydrodynamics in describing RHIC data
\cite{HiranoPhysRevC66} and by recent calculations of transport
coefficients of strongly interacting quark-gluon system using the
AdS/CFT correspondence \cite{KovtunPRL94}. The question is whether
dissipative hydrodynamics ($d$-hydrodynamics) is really needed and
how it should be implemented regarding problems with its
formulation like ambiguities in the form equations used
\cite{Tsumura2007}, unphysical instability of the equilibrium
state in the first order theory \cite{HiscockPhysRevD31} or loss
of causality in the first order equation approach
\cite{KoidePRC75}. We would like to propose a novel view on the
dissipative hydrodynamics based on nonextensive formulation of the
ideal hydrodynamical model, which we call the $q$-hydrodynamics
\cite{Osada2007}. When applied to ideal $q$-fluid it can be solved
exactly, in way similar to the usual ideal hydrodynamics. However,
it contains additional terms which can be interpreted as due to
dissipative effects expressed by the nonextensivity parameter $q$
- a single parameter here. Therefore, the following {\it
nonextensive/dissipative correspondence} (NexDC) emerges: ideal
$q$-fluid is apparently equivalent to some viscous fluid with its
transport coefficients being (implicit) functions of parameter
$q$. This parameter combines information about all possible
intrinsic fluctuations and correlations existing in the collision
process (in particular in the QGP being formed). Referring to
\cite{Osada2007} for more information on $q$-statistics it is
enough to say here that it is based on (indexed by $q$ and
nonextensive, see left panel of Fig. \ref{Fig12}) Tsallis rather
than Boltzmann-Gibbs (BG) entropy to which it converges for $q
\rightarrow 1$. Characteristic feature here is appearance of
$q$-exponentials, $\exp_q(-X) = [1-(1-q)X]^{1/(1-q)} \rightarrow
\exp(-X)$ for $q \rightarrow 1$.  Among other things $q-1$
measures scaled variance of the corresponding intensive quantities
like, for example, temperature $T$, or the amount of nonvanishing
in the hydrodynamical limit correlations (see left panel of Fig.
\ref{Fig12}) \cite{Osada2007}). Although $q$-hydrodynamics does
not fully solve the problems of $d$-hydrodynamics, nevertheless it
allows us to extend the usual perfect fluid approach (using only
one new parameter $q$) well behind its usual limits toward the
regions reserved for dissipative approach only.

In this note we can only explain main points of our proposition
leaving interested reader to \cite{Osada2007} and references there
for details. Our idea is visualized in Fig. \ref{Fig12}. Left
panel shows that in the usual hydrodynamics there is some spacial
scale $L_{hyd}$, such that volume $L_{hyd}^3$ contains enough
particles composing our fluid. However, in case when there are
some fluctuations and/or correlations in the system characterized
by some typical correlation length $l$ and when of $l> L_{hyd}$,
taking the usual limit $L_{hyd}\to 0$ removes the explicit
dependence on the scale $L_{hyd}$ but the correlation length $l$
leaves its imprint as parameter $q$ and one has to use
nonextensive entropy (one can argue that in this case $q \sim
l/L_{hyd}  \geq 1$). The situation encountered is shown on right
panel of Fig. \ref{Fig12}. Locally conserved BG entropy current
$s^{\mu}(x)$ and expensive entropy $S$ is replaced by locally
conserved $q$entropy current and nonextensive entropy $S_q$ local
equilibrium is replaced by a kind of stationary state (or
$q$-equilibrium containing dynamic leading to the assumed
intrinsic fluctuations/correlations and summarily characterized by
the parameter $q$).

The relativistic $q$-hydrodynamics is formulated \cite{Osada2007}
starting from the nonextensive Boltzmann equation
\cite{LavagnoPhysLettA301} leading to nonextensive entropy
($q$-entropy) current  \cite{Osada2007,LavagnoPhysLettA301}:
\begin{eqnarray}
 \sigma_q^{\mu}(x)  =
-k_{\rm B} \!\int \!\! \frac{d^3p}{(2\pi \hbar )^3}
\frac{p^{\mu}}{p^0} \Big\{ f_q^q \ln_q f_q  -f_q
\Big\},~~~~~~~~~~{\rm where}\qquad \ln_q f_q \equiv
[f_q^{(1-q)}-1]/(1-q) \label{eq:q_entropy}
\end{eqnarray}
where $f_q=f_q(x,p)$ is nonextensive version of phase space
distribution function in space-time position $x$ and momentum $p$.
One finds that $\partial_{\mu}\sigma_q^{\mu} \ge 0$ at any
space-time point, i.e., relativistic local $H$-theorem is valid
\cite{LavagnoPhysLettA301,Osada2007}. Demanding now that
$\partial_{\mu} \sigma_q^{\mu}\equiv 0$ one gets
\begin{eqnarray}
  f_q(x,p) \!\!&=&\!\! \left[ 1-(1-q)\frac{p_{\mu}u_{q}^{\mu}(x)}{k_{\rm B}
  T_q(x)}\right]^{1/(1-q)},
\label{eq:q-equilibrium}
\end{eqnarray}
where $T_q(x)$ is the $q$-temperature \cite{Osada2007} and
$u_{q}^{\mu}(x)$ is the $q$-hydrodynamical flow four-vector. The
state characterized by $f_q(x,p)$ is the {\it local
$q$-equilibrium state}, i.e.,  a kind of stationary state which
includes already some interactions between particles composing our
fluid (see right panel of Fig. \ref{Fig12}). The symmetry of
collision term in the nonextensive Boltzmann equation
\cite{Osada2007,LavagnoPhysLettA301} and energy-momentum
conservation in two particle collisions result in the nonextensive
version of local energy-momentum conservation,
\begin{equation}
T^{\mu\nu}_{q;\nu} = 0,\qquad {\rm where}\qquad {\cal
T}^{\mu\nu}_q(x) \equiv \frac{1}{(2\pi\hbar)^3}\int
\frac{d^3p}{p^0} p^{\mu}p^{\nu} f_q^q(x,p).
\label{eq:q-energy_momentum_tensor}
\end{equation}
Assuming now that this $q$-energy-momentum tensor can be
decomposed in the usual way in terms of the $q$-modified energy
density $\varepsilon_q$ and $q$-pressure $P_q$ by using the
$q$-modified flow $u_q^{\mu}$ (such that $u_q^{\mu}=(1,0,0,0)$ in
the rest frame) one obtains the {\it perfect $q$-hydrodynamical
equation} ($\Delta_q^{\mu\nu}\equiv g^{\mu\nu}-
u_q^{\mu}u_q^{\nu}$) (covariant derivative notation is used here,
 see \cite{Osada2007}):
\begin{eqnarray}
{\cal T}_{q;\mu}^{\mu\nu} = \Big[ \varepsilon_q (T_q)
u_q^{\mu}u_q^{\nu} - P_q (T_q)\Delta_q^{\mu\nu} \Big]_{;\mu} =0.
\label{eq:q-equation_of_motion}
\end{eqnarray}
One the other hand one can also decompose ${\cal T}_q^{\mu\nu}$,
using the usual $4$-velocity fluid field $u^{\mu}$ and obtain
equation
\begin{eqnarray}
\left[\tilde{ \varepsilon}(T_q) u^{\mu}u^{\nu}
 -\tilde{P}(T_q)\Delta^{\mu\nu}
 + 2W^{(\mu}u^{\nu)}
 +\pi^{\mu\nu} \right]_{;\mu} \!\!\!=0 ,
\label{eq:decomposition}
\end{eqnarray}
where ($\delta u_q^{\mu} \equiv u_q^{\mu}-u^{\mu}$ and
$\Delta^{\mu\nu} \equiv g^{\mu\nu}- u^{\mu}u^{\nu}$) whereas
\begin{eqnarray}
 &&\tilde{\varepsilon}=\varepsilon_q+3\Pi ; \qquad \qquad \tilde{P} = P_q +\Pi;
 \qquad {W}^{\mu}  = w_q[1+x] ~\Delta^{\mu}_{\lambda} \delta u_q^{\lambda}; \nonumber\\
 &&{\pi}^{\mu\nu} = \frac{W^{\mu} W^{\nu}}{ w_q [1+x] ^2} +\Pi\Delta^{\mu\nu}
 =w_q~ \delta u_q^{<\mu}  \delta u_q^{\nu >};\qquad
 \Pi  \equiv \frac{1}{3} w_q [x^2+2x]
\end{eqnarray}
are, respectively, energy density ($\tilde{\varepsilon}$),
pressure ($\tilde{P}$), energy or heat flow vector ($W^{\mu}$),
shear (symmetric and traceless) pressure tensor ($\pi^{\mu\nu}$)
and bulk pressure ($\Pi$). Notation used is: $A^{( \mu}B^{\nu
)}\equiv \frac{1}{2}(A^{\mu}B^{\nu} + A^{\nu}B^{\mu});\qquad w_q
\equiv \varepsilon_q + P_q;\qquad  x \equiv u_{\mu}\delta
u_q^{\mu} = -\frac{1}{2}\delta u_{q\mu} \delta u_q^{\mu};\qquad
a^{<\mu}b^{ \nu >} \equiv [\frac{1}{2}( \Delta^{\mu}_{\lambda}
\Delta^{\nu}_{\sigma} + \Delta^{\mu}_{\sigma}
\Delta^{\nu}_{\lambda} ) -\frac{1}{3}\Delta^{\mu\nu}
\Delta_{\lambda\sigma} ] a^{\lambda}b^{\sigma}$.  Notice that
whereas the time evolution of $\Pi$ is controlled by
$q$-hydrodynamics (via the respective time dependencies of
$\varepsilon_q$, $P_q$ and $x$) its form is determined by the
assumed constraints which must assure that the local entropy
production in the standard $2^{nd}$ order theory
\cite{IsraelAnnPhys118,MurongaPRC69} is never negative.

The crucial point of our work is assumption that there exists some
temperature $T$ and velocity $\delta u_q^{\mu}$ satisfying the
following NexDC relations:
\begin{eqnarray}
P(T)= P_q(T_q),\quad \varepsilon(T)= \varepsilon_q(T_q) +3\Pi
\label{eq:Nex/diss}
\end{eqnarray}
($\varepsilon$ and $P$ are energy density and pressure defined in
the usual Boltzmann-Gibbs statistics, i.e., for $q=1$).In this
case one can transform equation (\ref{eq:decomposition}) into the
following usual $d$-hydrodynamical equation \cite{Osada2007}:
\begin{eqnarray}
\left[ \varepsilon(T) u^{\mu}u^{\nu} \!-\!(P(T)+\Pi )
\Delta^{\mu\nu} \!\!\!+\! 2 W^{(\mu} u^{\nu )} \!+\!\pi^{\mu\nu}
\right]_{;\mu} \!\!\!\!=0. \label{eq:Nex/diss_equation}
\end{eqnarray}
This completes demonstration of the equivalence of perfect
$q$-hydrodynamics represented by Eq.
(\ref{eq:q-equation_of_motion}) and its $d$-hydrodynamics
counterpart represented by Eq. (\ref{eq:Nex/diss_equation}). We
propose to call this equivalence NexDC: {\it
nonextensive/dissipative correspondence}. In \cite{Osada2007} we
have successfully applied $q$-hydrodynamics to description of RHIC
data on particle production (in a limited fashion, however; to
apply it to flow effects and correlation phenomena, like HBT
effect, one must go out the one-dimensional approximation used
here - such work is now in progress).

The most important point in in NexDC is the fact that although in
ideal $q$-hydrodynamics the $q$-entropy is {\it conserved}, i.e.,
$[ s_q u_q^{\mu}]_{;\mu}=0$, we can rewrite it in the form
corresponding to dissipative fluid with {\it entropy production}:
$[s u^{\mu} ]_{;\mu} =-\frac{u_{\nu}}{T}\delta {\cal
 T}^{\mu\nu}_{;\mu}$. When applied to description of multiparticle
 production processes this fact is seen in the prediction of $q$-dependent
 increase of multiplicity of produced particles \cite{Osada2007}.
 The most general expression for the full order dissipative entropy current
in the NexDC approach:
\begin{eqnarray}
\sigma^{\mu}_{\rm full} \!\!&\equiv&\!\!
 su^{\mu} +\frac{W^{\mu}}{T} -\frac{2T}{T_q}\left[~1- \sqrt{1-\frac{3\Pi}{w}}
~\right]su^{\mu} + \frac{2(T-T_q)}{T_q}\frac{W^{\mu}}{T}.
\label{eq:full_order_current}
\end{eqnarray}
\begin{figure}[h!]
\begin{center}
\includegraphics[width=7.4cm]{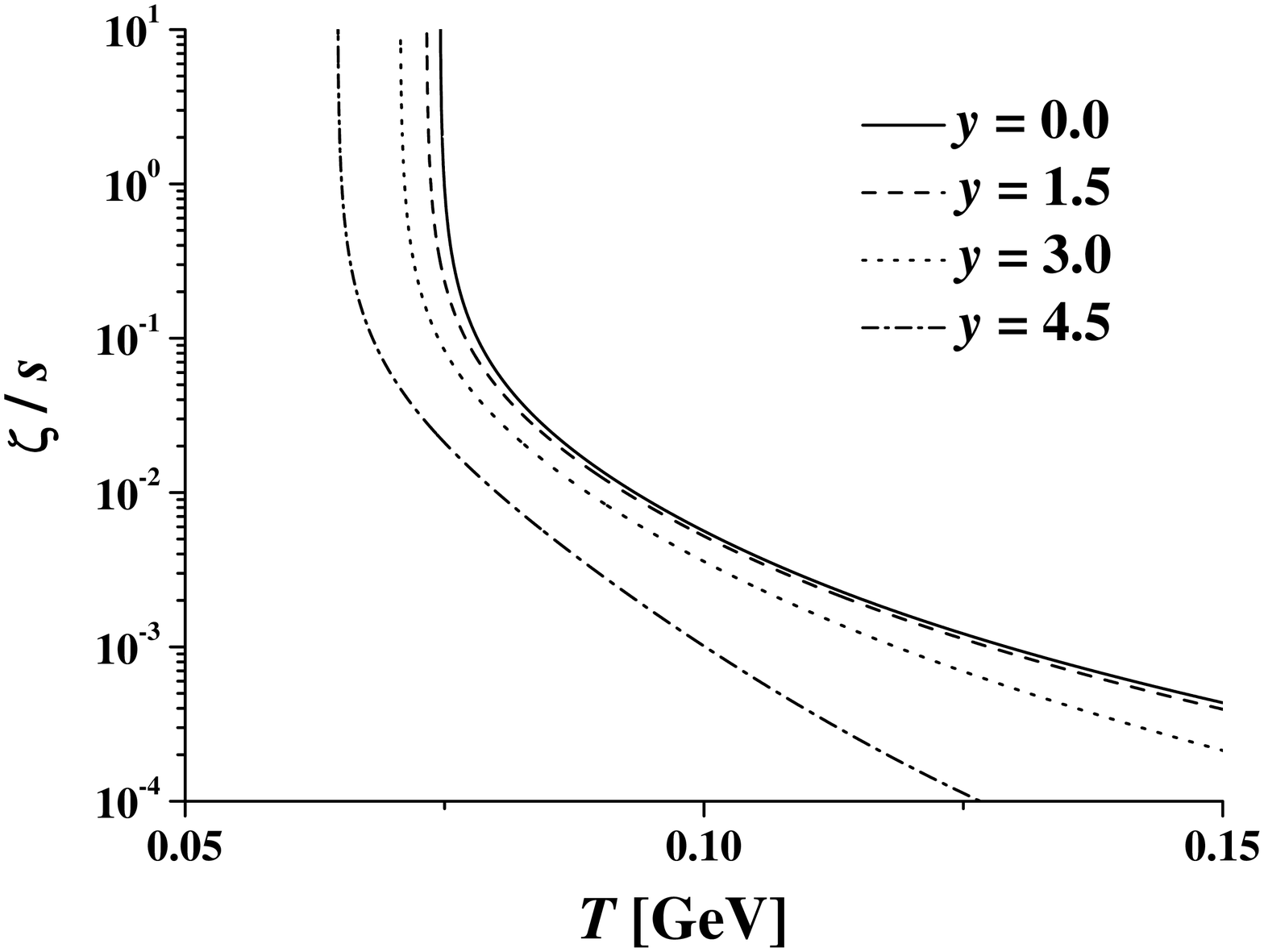}\hfill
\includegraphics[width=7.4cm]{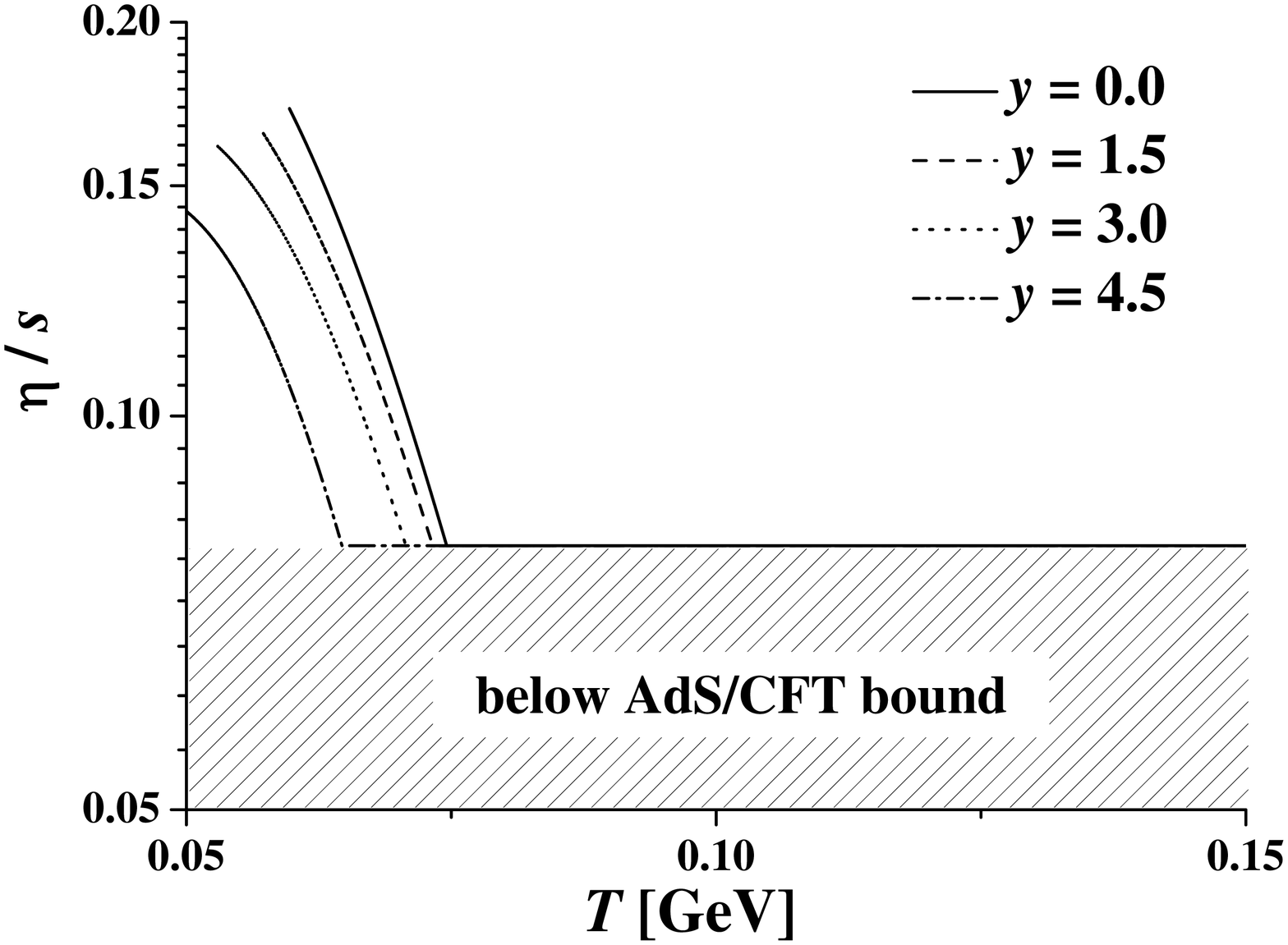}
\caption{The NexDC predictions for the ratios of bulk (left panel)
and shear (right panel) viscosities over the entropy density,
$\zeta/s$ and $\eta/s$, as function of temperature $T$ calculated
for a number of space-time rapidities $y\equiv
\frac{1}{2}\ln\frac{t+z}{t-z}$ using $q$-hydrodynamical model
developed in \cite{Osada2007} with $q=1.08$. The $q$-initial
conditions with initial energy density $\varepsilon^{(in)}=22.3$
GeV/fm$^3$ and $q$-equation of state for relativistic $\pi$ gas
was used (see \cite{Osada2007} for details).}\label{Fig12ab}
\end{center}
\end{figure}

One can now calculate bulk and shear viscosities emerging from the
NexDC. Here we shall present only our main result, namely the sum
rule connecting ratios of bulk and shear viscosities over the
entropy density $s$:
\begin{eqnarray}
 \frac{1}{\zeta/s}+\frac{3}{\eta/s} = \frac{w \sigma^{\mu}_{{\rm
 full};\mu}}{\Pi^2}.
\label{eq:sum_rule}
\end{eqnarray}
To disentangle it some additional input is needed. Results
presented in Fig. \ref{Fig12ab} are obtained assuming that total
entropy is generated by action of the shear viscosity only. This
can be confronted with Ads/CFT conjecture \cite{KovtunPRL94} that
$\eta/s \ge 1/4\pi$.

To summarize: we claim that nonextensive approach to hydrodynamics
can be regarded as a new phenomenological way to deal with viscous
fluids in which many different dynamical features (already known
or yet to be discovered) are summarily represented by a single
parameter $q$ describing a kind of $q$-ideal fluid by means of
ideal $q$-hydrodynamics, which is apparently much simpler to
handle than the usual dissipative hydrodynamics.\\

Partial support (GW) of the Ministry of Science and Higher
Education under contracts 1P03B02230 and  CERN/88/2006 is
acknowledged.\\


\begin{thebibliography}{99}

\bibitem{IsraelAnnPhys118} W. Israel, Ann. Phys. (N.Y.) 100 (1976) 310; J.M. Stewart,
                           Proc. R. Soc. London A357 (1977) 59;
                           W. Israel and J.M. Stewart, Ann~Phys. (N.Y.) 118 (1979)341.

\bibitem{MurongaPRC69} A. Muronga, Phys. Rev. C69 (2004) 034903
                       and Phys. Rev. Lett. 88 (2002) 062302;
                       A. Muronga and D.H. Rischke, arXiv:nucl-th/0407114;
                       H. Song and U. Heinz, arXiv:0712.3715 and references therein;
                       A. Dumitru, E. Moln\'ar and Y. Nara, Phys. Rev. C76 (2007)
                       024910;
                       P. Romatschke and U. Romatschke, Phys. Rev.
                       Lett. 99 (2007) 17230 and references therein.

\bibitem{Tsumura2007} K. Tsumura and T. Kunihiro, arXiv:0709.3645 and references therein.

\bibitem{HiscockPhysRevD31} W.A. Hiscock and L. Lindblom, Ann.~Phys. (N.Y.) 151 (1983) 466;
                            Phys. Rev. D31 (1985) 725 and D35 (1987) 3723.

\bibitem{KoidePRC75} T. Koide, G.S. Denicol, Ph. Mota and T. Kodama, Phys. Rev.
                     C75 (2007) 034909.

\bibitem{HiranoPhysRevC66} Cf., for example, T. Hirano and K. Tsuda, Phys. Rev.
                           C66 (2002) 054905 (2002) or
                           D. Teaney, Phys. Rev. C68 (2003) 034913.

\bibitem{KovtunPRL94} P. Kovtun, D.T. Son and A.O. Starinets, Phys. Rev. Lett.
                      94 (2005) 111601. For most recent reviews of this subject
                      see D.T. Son ans A.O. Starinet, Ann. Rev. Nucl. Part. Sci. 57 (2007) 95
                      and references therein.

\bibitem{Osada2007} T. Osada and G. Wilk, Phys. Rev. C77 (2008) 044903 [arXiv:0710.1905].

\bibitem{LavagnoPhysLettA301} A. Lavagno, Phys. Lett. A301 (2002) 13.





\end{thebibliography}
\end{document}